\documentclass{article}
\usepackage{stywhispers,amsmath,epsfig, subcaption, comment}
\usepackage{tikz}
\usepackage[absolute,overlay]{textpos}
\usepackage{blindtext}
\usepackage{relsize}
\usepackage{hyperref}

\tikzset{fontscale/.style={font=\relsize{#1}}}

\title{Reliable Explainability of Deep Learning Spatial-spectral Classifiers for Improved Semantic Segmentation in Autonomous Driving}

\name{Jon Gutiérrez-Zaballa$^1$\sthanks{This work was partially supported by the University of the Basque Country (UPV-EHU) under grant GIU21/007, by the Basque Government under grants PRE\_2023\_2\_0148 and KK-2023/00090, and by the Spanish Ministry of Science and Innovation under grant PID2020-115375RB-I00.}, Koldo Basterretxea$^1$, Javier Echanobe$^2$}
\address{Department of Electronics Technology, University of the Basque Country, Bilbao, Spain$^1$ \\
Department of Electricity and Electronics, University of the Basque Country, Leioa, Spain$^2$}

\begin{document}
%
\maketitle

\begin{textblock*}{21cm}(1.5cm,26cm)
\begin{tikzpicture}
    \draw (0,0) rectangle (17.5,0.5); 
    \end{tikzpicture}
\end{textblock*} 

\begin{textblock*}{21cm}(0cm,26cm)
\begin{tikzpicture}
    \node (center) {c};
    \path (center)+(10.5,4) node [fontscale=-1] (name) {\copyright 2024 IEEE. Final published version of the article can be found at \href{https://ieeexplore.ieee.org/document/10876465}{10.1109/WHISPERS65427.2024.10876465}.};
    \end{tikzpicture}
\end{textblock*} 

\begin{abstract}
Integrating hyperspectral imagery (HSI) with deep neural networks (DNNs) can strengthen the accuracy of intelligent vision systems by combining spectral and spatial information, which is useful for tasks like semantic segmentation in autonomous driving.
To advance research in such safety-critical systems, determining the precise contribution of spectral information to complex DNNs' output is needed.
To address this, several saliency methods, such as class activation maps (CAM), have been proposed primarily for image classification.
However, recent studies have raised concerns regarding their reliability.
In this paper, we address their limitations and propose an alternative approach by leveraging the data provided by activations and weights from relevant DNN layers to better capture the relationship between input features and predictions.
The study aims to assess the superior performance of HSI compared to 3-channel and single-channel DNNs.
We also address the influence of spectral signature normalization for enhancing DNN robustness in real-world driving conditions.
\end{abstract}


\section{Introduction}
A key challenge in computer vision is metamerism, which hinders the robustness of tristimulus based systems.
Metamerism occurs when the reflectance curves of different materials match under certain conditions, like specific illumination or sensor characteristics, but differ under others \cite{wright1944measurement}.

Hyperspectral sensors offer a promising solution to address metamerism and increase the robustness of intelligent vision systems.
These sensors capture light across narrow-band filters usually spanning visible to infrared wavelengths.
Hyperspectral imaging (HSI) has proven valuable in fields such as food quality assessment, medical analysis, and remote sensing \cite{BHARGAVA2024e33208} and its extension to new application domains, such as autonomous driving systems (ADS), is currently being investigated.
Despite its benefits, HSI poses challenges for classification tasks due to its high dimensionality, inter-class similarity, and intra-class variability \cite{paoletti2019deep}.
To achieve pixel-level precision in applications like semantic or instance segmentation, combining spectral and spatial information is crucial.
One approach to achieving this is by using fully convolutional networks (FCNs), a type of deep neural network (DNN) with an encoder-decoder structure.

To comprehend the contribution of spectral information beyond the numerical evaluations provided by ablation experiments, DNNs must be interpretable, meaning we should be able to interpret their behavior in the decision-making processes.
Explainable AI, which focuses on the interpretability of neural networks, is a hot topic nowadays \cite{ali2023explainable}.
In the intelligent vision area, the most common methods for identifying which image features are most relevant for predictions are the saliency methods, including layer-wise relevance propagation (LRP), gradient-based techniques, class activation mapping (CAM), and their variants.
However, despite recent advancements, many of these methods exhibit notable limitations.
As noted by \cite{adebayo2018sanity, kindermans2019reliability, dieter2023evaluation, kim2023extending}, most saliency methods for image classification are designed to produce visually appealing representations rather than accurately reflecting the underlying prediction processes.

In this article, in Section \ref{sec:background}, we first review the desirable properties that saliency methods should have to be considered reliable, as discussed in the literature.
In Section \ref{sec:conservativenessDefinition}, we extend the CAM conservativeness property --originally defined for classification networks-- to semantic segmentation networks.
Next, in Section \ref{sec:conservativenessEvaluation}, we examine the essential characteristics a CAM-based saliency method for segmentation must have to satisfy this property, demonstrating that many such methods do not adequately fulfill it.
To overcome these limitations when interpreting the inference process on FCNs, in Section \ref{sec:spectralRichness} we propose a method  that leverages activations and weights from specific, relevant model layers to more reliably explain the relationship of input information with predictions.
We apply this method to evaluate a segmentation U-Net trained with images from a 25-band hyperspectral snapshot camera in real-world driving scenarios, assessing how spectral information contributes to the enhanced performance of HSI-based classification models.
Finally, Section \ref{sec:discussion} provides a discussion of our findings and their implications.
It also outlines directions for future work.

\section{Background}\label{sec:background}
\subsection{Invariance Tests for Saliency Methods}
The reliability of saliency methods has recently been questioned, prompting efforts to identify the invariants that saliency methods should satisfy to be considered trustworthy.
As highlighted by the authors in \cite{adebayo2018sanity}, some widely used techniques are independent of model parameters and training data, thus failing to establish valid relationships between inputs and outputs present in the data.

In this context, \cite{adebayo2018sanity} conducts two sanity checks on various saliency methods: one based on model parameter randomization (comparison against a randomly initialized untrained network) and the other on data randomization (comparison against a trained network but with randomly permuted labels).
GradCAM passes these sanity checks, while Guided BackProp and GuidedGradCAM are invariant to parameters in higher layers, hence failing.
However, it is important to note that when dealing with overparameterized, non-optimized models (unpruned/unquantized models), certain layers in the network may contribute minimally to the output, leading to some level of invariability to parameter perturbations.
Similarly, the authors of \cite{dieter2023evaluation} show, through adversarial examples, that LRP fails to explain the DNN’s decision-making process for original images or adversarial examples, posing a further challenge for reliable explainability methods.

The authors of \cite{kindermans2019reliability} propose the requirement of input invariance to ensure a reliable interpretation of the input’s contribution to the model predictions.
This principle dictates that a saliency method should reflect the model's sensitivity to transformations of the inputs.
Finally, in \cite{kim2023extending}, the authors introduce the concept of conservativeness in CAM.
Conservativeness implies that the sum of the contributions in the mapping aligns with the prediction scores.
The authors show that the original CAM \cite{zhou2016learning} satisfies this property, while GradCAM \cite{selvaraju2017grad} does not, leading to the proposal of a variant method called Extended-CAM.

\subsection{Basics of CAM-based Studies}\label{sec:camBasics}
All methods based on class activation mappings derive from the original CAM \cite{zhou2016learning}.
The goal is to investigate the contribution of the spatial component $(i,j)$ to the final prediction $y^c$ for an image classified as belonging to class $c$.
As previously mentioned, the sum of these contributions ($L_{ij}^c$) should match $y^c$ to satisfy the conservativeness property (Eq. \ref{equ:conservativeness}).

\begin{equation}
    \sum_{i, j} L_{i, j}^c = y^c
    \label{equ:conservativeness}
\end{equation}

However, the original CAM is limited by the assumption that the last layer of the feature extractor consists of a global average pooling layer followed by a fully-connected layer, which restricts its applicability to modern DNNs.

To extend the use of CAM to more generic classification networks, GradCAM (Eq. \ref{equ:gradCAM}) was introduced in \cite{selvaraju2017grad}:

\begin{equation}
    L_{i,j}^{c} = ReLU(\sum_{k} A_{ijk} \ \alpha_{k}^{c})
    \label{equ:gradCAM}
\end{equation}

where $A_{ijk}$ is the activation of a certain layer and $\alpha_{k}^{c}$ is the linear multiplying coefficient.
Alternative methods based on CAM propose different ways to calculate these $\alpha_{k}^{c}$ coefficients.
In the original GradCAM, these coefficients are determined through a global average pooling operation (Eq. \ref{equ:gradCAMcoeff}):

\begin{equation}
    \alpha_{k}^{c} = \frac{1}{Z} \sum_{u} \sum_{v} \frac{\partial y^c}{\partial A^k_{u,v}}
    \label{equ:gradCAMcoeff}
\end{equation}

The corresponding approach for segmentation tasks, SegGradCAM, was proposed in \cite{Vinogradova_Dibrov_Myers_2020} and is detailed in (Eq. \ref{equ:segGradCAM}).

\begin{equation}
    L_{i,j}^c = ReLU(\sum_{k} A_{ijk} \frac{1}{N} \sum_{u,v} \frac{\partial \sum_{(r,s) \in M} y_{rs}^c }{\partial A_{uv}^k})
    \label{equ:segGradCAM}
\end{equation}

where $M$ is the set of pixels in the region of interest (RoI), and $y^c_{rs}$ is the non-probabilistic output (logit) for the pixel at $(r,s)$ belonging to class $c$.
When $M$ contains only one pixel, SegGradCAM (Eq. \ref{equ:segGradCAM}) simplifies to GradCAM (Eq. \ref{equ:gradCAM}), even though they describe different input/output connections.

\section{Extending conservativeness to semantic segmentation}\label{sec:conservativenessDefinition}
Conservativeness ensures the precise quantitative explanation of prediction scores by providing absolute pixelwise contributions without redundancy or deficiency \cite{kim2023extending}.
This property is crucial for visualization methods in neural networks focused on semantic segmentation, as it allows for analyzing the input contributions to a single output pixel, specific pixels in a RoI, or every pixel associated with a certain label.
If this condition is not met, the validity of comparing CAMs from different layers or DNNs becomes questionable.

In \cite{kim2023extending}, conservativeness is proposed for image classification models.
We extend this property to semantic segmentation models in Eq. \ref{equ:conservativenessSegmentation} which indicates that a pixel activation mapping $L_{i,j,r,s}^c$ for class $c$ and output location $(r,s)$ is considered conservative if and only if the spatial sum over $(i,j)$ equals the target score $y_{r,s}^c$ for every pixel.

\begin{equation}
    \sum_{i,j} L_{i,j,r,s}^c = y_{r,s}^c
    \label{equ:conservativenessSegmentation}
\end{equation}

Class-level conservativeness is defined in Eq. \ref{equ:conservativenessSegmentation2} and depends on pixel-level conservativeness and vice versa.

\begin{equation}
    \sum_{r,s \in M} (y^c_{r,s} \cdot m_{r,s}) = \sum_{r,s \in M}[(\sum_{i,j} L_{i,j,r,s}^c ) \cdot m_{r,s}] = Y^c
    \label{equ:conservativenessSegmentation2}
\end{equation}

where $y^c_{r,s}$ is the logit for the pixel at position $(r,s)$ belonging to class $c$, $m(r,s)$ is a 0-valued matrix with the same size as the output, containing a 1 at $(r,s)$ and $Y^c$ is the logit map for class $c$.
Finally, the segmentation class activation mapping is defined as in Eq. \ref{equ:conservativenessSegmentation3}.

\begin{equation}
    L_{i,j}^c = \sum_{r,s \in M} L_{i,j,r,s}^c
    \label{equ:conservativenessSegmentation3}
\end{equation}

\begin{figure}[h!]
\centering
\includegraphics[width=8.5cm]{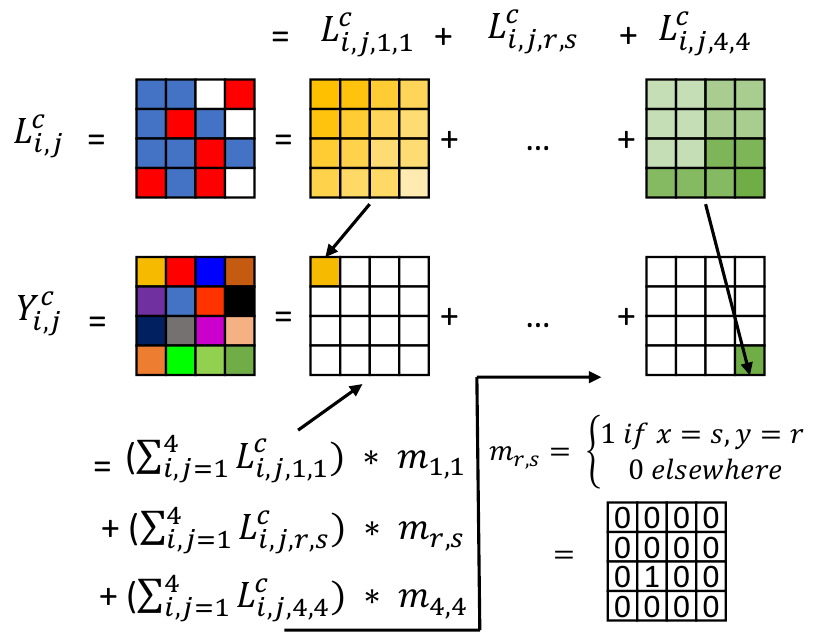}
\caption{Conservativeness property in semantic segmentation.}
\label{fig:conservativenessSegmentation}
\end{figure}

The interpretation of Eqs. \ref{equ:conservativenessSegmentation}, \ref{equ:conservativenessSegmentation2} and \ref{equ:conservativenessSegmentation3} is illustrated in Fig. \ref{fig:conservativenessSegmentation}.
Here, $L_{i,j}^c$ is the segmentation activation mapping (with red for positive contributions, blue for negative ones and white for no contributions) for class $c$, created by adding the pixel activation maps for all pixels in the RoI.
For conservativeness to hold, the sum of the elements in each pixel activation map must match the logit value for that pixel.
The spatial arrangement of these logits forms the output logit map, $Y^c$, for the pixels within the RoI.

\section{Conservativeness assessment in GradCAM- and SegGradCAM-based methods}\label{sec:conservativenessEvaluation}
As previously discussed in Section \ref{sec:background}, many saliency methods do not adhere to fundamental invariants.
Hereafter, we show that several widely used CAM-based methods, included some designed for classification, also fail to meet the conservativeness principle.
The selection of these specific methods is motivated by their prominence in the literature and their unique approaches for calculating $\alpha_{k}^{c}$ coefficients (Eq. \ref{equ:gradCAMcoeff}).

To illustrate the non-conservativeness of gradient-based saliency methods, we just can rewrite Equation \ref{equ:conservativeness} in the form of GradCAM (Eq. \ref{equ:gradCAM}), which is known not to meet the conservativeness criterion due to the spatial-averaging global pooling operation performed (Eq. \ref{equ:gradCAMcoeff}), as noted in \cite{kim2023extending}.

SegGradCAM (see Subsection \ref{sec:camBasics}) modifies Eq. \ref{equ:segGradCAM} into Eq. \ref{equ:segGradCAMRewritten} by applying the commutative property between partial derivatives and summation, while omitting the ReLU operation for simplicity.

\begin{equation}
L_{i,j}^c = \sum_{(r,s) \in M} L_{i,j,r,s}^c = \sum_{(r,s) \in M} \sum_{k} A_{ijk} (\frac{1}{N} \sum_{u,v} \frac{\partial y_{rs}^c}{\partial A_{uvk}})
\label{equ:segGradCAMRewritten}
\end{equation}

\begin{figure*}[h!]
  \begin{subfigure}{0.33\textwidth}
    \includegraphics[width=5.75cm]{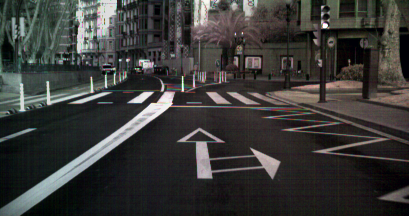}
    \caption{}
    \label{fig:exampleImage}
  \end{subfigure}
  \hfill
  \begin{subfigure}{0.33\textwidth}
    \includegraphics[width=5.75cm]{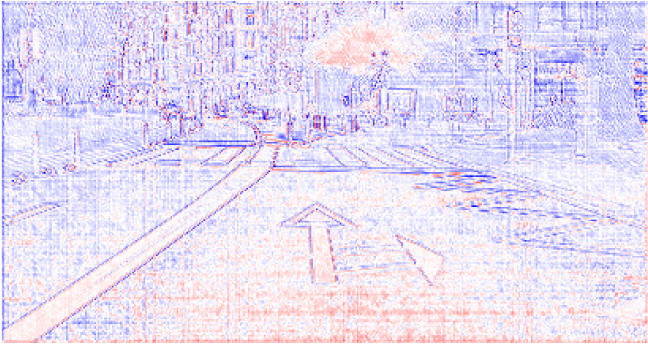}
    \caption{}
    \label{fig:originalSegGradCAM}
  \end{subfigure}%
  \hfill
  \begin{subfigure}{0.33\textwidth}
    \includegraphics[width=5.75cm]{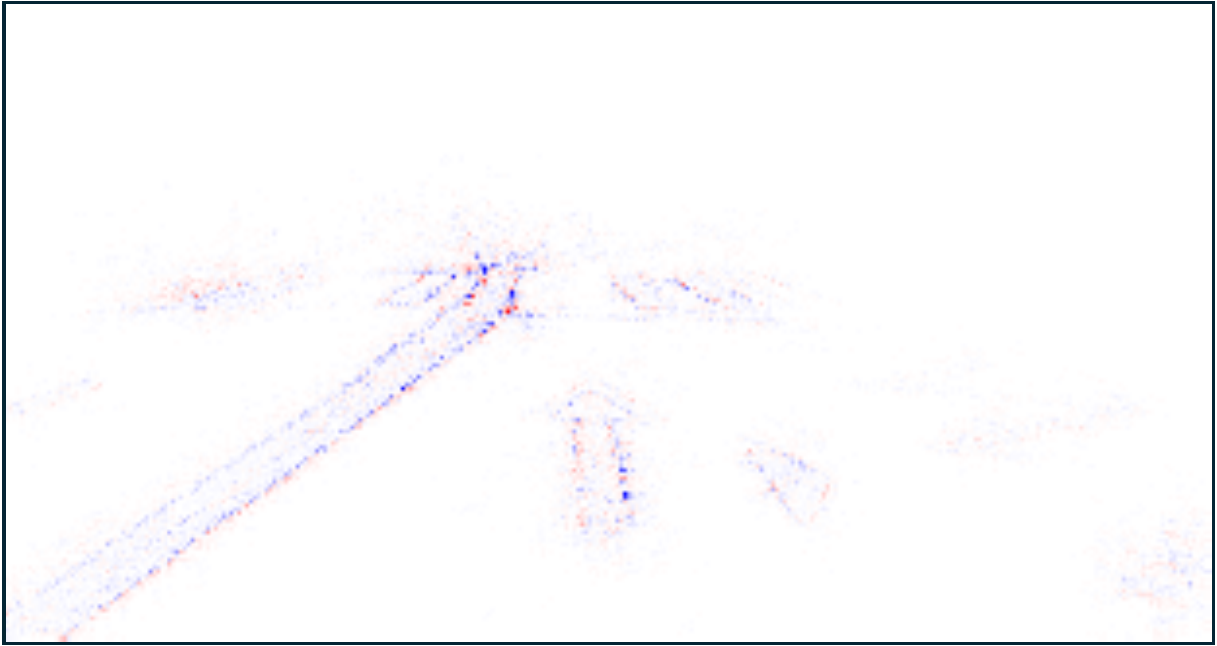}
    \caption{}
    \label{fig:modifiedSegGradCAM}
  \end{subfigure}%
  \caption{Effect of the SegGradCAM gradient weighting modification on $Conv2D\_1$ activation from HSI \& PN model.
  Image 1111\_576 from HSI-Drive v2.0 \cite{HSIDriveV20}.
  a) Pseudocolor version.
  b) Spatial average.
  c) Non-spatial average.}
\end{figure*}

When $M$ contains multiple pixels, each pixel has a specific logit activation map $L_{i,j,r,s}^c$, which is added to obtain the logit map of class $c$, as illustrated in Fig. \ref{fig:conservativenessSegmentation}.
For a single pixel in $M$, the equations for SegGradCAM and GradCAM are essentially identical, making the demonstration of non-conservativeness straightforward.

The use of global average pooling over the gradient in calculating $\alpha_{k}^{c}$ for semantic segmentation introduces unwanted effects.
For instance, using an image from HSI-Drive \cite{HSIDriveV20}, the CAM for the Road Marks class (Fig. \ref{fig:originalSegGradCAM}) shows nearly every pixel contributing, which is counterintuitive.
To address this, we can eliminate the average from $\alpha_{ijk}^{c}$ calculation, giving spatial dimensions to it.
The equation for SegGradCAM with the corrected $\alpha_{ijk}^{c}$ is shown in Eq. \ref{equ:extendedSegGradCAM}.

\begin{equation}
    L_{i,j}^{c} = \sum_{(r,s) \in M} L_{i,j,r,s}^c = \sum_{(r,s) \in M} (\sum_{k} A_{ijk} \odot \frac{\partial y_{rs} } {\partial A_{ijk}})
    \label{equ:extendedSegGradCAM}
\end{equation}

This correction yields a more meaningful class activation mapping for the Road Marks class, as shown in Fig. \ref{fig:modifiedSegGradCAM}.
However, a more reasonable appearance does not ensure adherence to the conservativeness principle.
Our evaluation demonstrates that Eq. \ref{equ:extendedSegGradCAM} can also be derived by extending Extended-CAM \cite{kim2023extending} formulation, which satisfies conservativeness principle for classification, to the segmentation task.

The authors of ExtendedCAM suggest clipping negative values of the class activation mapping using ReLU functions, claiming it enhances visualization accuracy.
However, this practice violates the conservativeness principle and risks losing critical information regarding regions contributing negatively to the classification.
Furthermore, for ExtendedCAM and its extension to segmentation to maintain conservativeness, nonlinear layers of the network must be approximable by the first term of a Taylor decomposition.
This is a condition not generally applicable to segmentation DNNs employing standard data processing structures, including Max-Pooling and Concatenation layers like those in U-Net.

The SegXResCAM method proposed in \cite{hasany2023seg} employs pooling/unpooling operations on gradients to regulate their granularity.
Nevertheless, similar to the average pooling in GradCAM, this spatial processing compromises conservativeness.
The classification non-conservativeness of SegXResCAM is evidenced in Eq. \ref{equ:segXResCAMconservativeness}.

\begin{equation}
    \sum_{i,j} L_{i,j}^c = \sum_{i,j} ReLU(\sum_{k} Up[Pool[\frac{\partial Y^c}{\partial A_{ijk}}]] \odot A_{ijk}) \ne y^c
    \label{equ:segXResCAMconservativeness}
\end{equation}

In what follows, we also evaluate RelevanceCAM \cite{Lee_2021_CVPR} and ScoreCAM \cite{ScoreCAM}, which are not gradient-based but also fail to satisfy conservativeness.

RelevanceCAM (Eq. \ref{equ:RelevanceCAMconservativeness}) does not meet conservativeness because it derives weights from a global average pooling operation over the relevance maps $R^{(c, i)} (x,y)$.
\begin{equation}
    \sum_{i,j} L_{i,j}^c = \sum_{i,j} \sum_{k} (\sum_{x, y} R_k^{(c, i)} (x,y)) A_k^c \ne y^c
    \label{equ:RelevanceCAMconservativeness}
\end{equation}

ScoreCAM (Eq. \ref{equ:ScoreCAMconservativeness}) defines the weighting of activation maps $C(A_l^k)$ through a channel-wise increase of confidence score based on perturbation analysis.
Its definition, differing from the network's inference process, results in non-conservativeness.

\begin{equation}
    \sum_{i,j} L_{i,j}^c = \sum_{i,j} ReLU(\sum_{k} C(A_l^k) A_l^k) \ne y^c
    \label{equ:ScoreCAMconservativeness}
\end{equation}

The absence of conservativeness across these methods impedes performing meaningful comparative analysis, particularly when evaluating CAMs across multiple models in our study.
While these techniques can still be valuable to understand model behavior, particularly for simpler classification tasks, this article emphasizes that saliency methods should focus on representing reality as accurately as possible and produce insightful heat maps rather than fancy, superficial visualizations.

\begin{figure*}[h!]
  \begin{subfigure}{0.24\textwidth}
    \includegraphics[width=\linewidth]{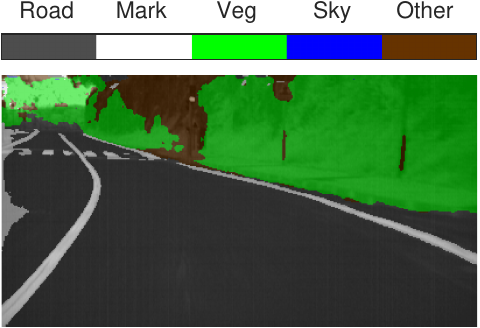}
    \caption{1-channel.}
    \label{fig:1channelSeg}
  \end{subfigure}%
  \hfill
  \begin{subfigure}{0.24\textwidth}
    \includegraphics[width=\linewidth]{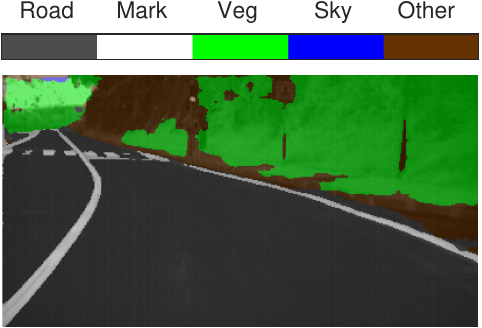}
    \caption{3-channel.}
    \label{fig:3channelSeg}
  \end{subfigure}%
  \hfill
  \begin{subfigure}{0.24\textwidth}
    \includegraphics[width=\linewidth]{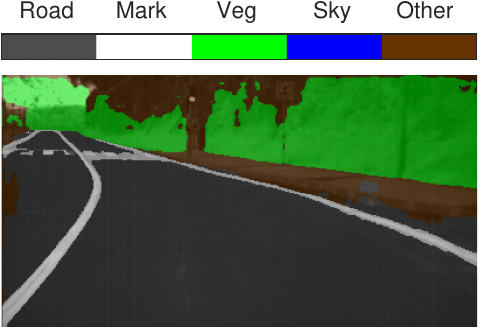}
    \caption{25-channel with PN.}
    \label{fig:25channelSeg}
  \end{subfigure}%
  \hfill
  \begin{subfigure}{0.24\textwidth}
  \vspace{7mm}
  \includegraphics[width=\linewidth]{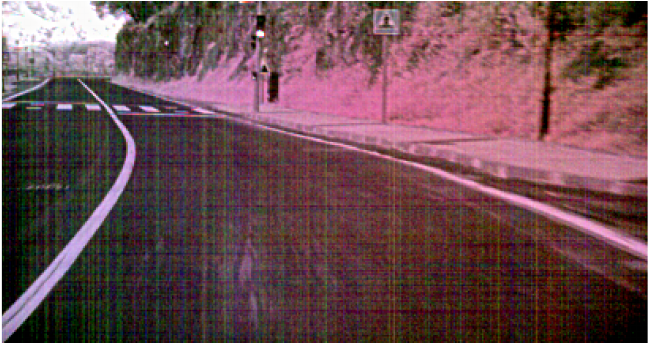}
  \caption{Pseudocolor 3112\_104.}
  \label{fig:pseudocolor104}
  \end{subfigure}
  \caption{Inference result on image 3112\_104 from HSI-Drive \cite{HSIDriveV20}.}
  \label{fig:SegmentationExample}
\end{figure*}

\section{Analyzing the contribution of spectral information in HSI segmentation DNN models}\label{sec:spectralRichness}
\subsection{A DNN for Scene Understanding Using HSI}
HSI-Drive \cite{HSIDriveV20} is a comprehensive dataset designed specifically for researching HSI technology in developing more accurate and robust ADS, particularly under adverse environmental conditions.
We trained and optimized various U-Net-based machine learning models on HSI-Drive using different classification targets to create image segmentation systems of varying complexity.
To assess the contribution of spectral information to the inference processes of these models, we evaluated the performance of the same model architecture using 1, 3, and all the 25 spectral channels provided by an Imec 5x5 mosaic-type near-infrared (NIR) HSI sensor.
Fig. \ref{fig:SegmentationExample} shows an example of the 5-class segmentation experiment comprising Road (tarmac), Road Marks, Vegetation, Sky, and "Other" classes.
For more in-depth information regarding the DNN architecture and the dataset used in this article, the reader is referred to \cite{HSIDriveV20, gutierrez2024evaluating}.

HSI-Drive was recorded with a snapshot camera equipped with a multispectral filter array (MSFA) based on Fabry-Perot interferometers.
This technology, when used with no rejection filters, as was the case, can lead to some input channels containing mixed contributions from distinct wavelengths \cite{HSIDriveV20}.
For this study, 1-channel images consist of the MSFA central band, which exhibits a spectral response centered around 731.883 nm.
3-channel images, which were generated as an alternative to RGB imaging (no visible RGB bands can be extracted from these images), consist of the three most informative spectral bands, which were extracted using the orthogonal space projection method as explained in \cite{GUTIERREZZABALLA2023102878}.
Specifically, these bands include: 9 (with peak at 770.576 nm), 22 (featuring a primary peak at 858.948 nm and a secondary peak at 577.147 nm), and 25 (having a primary peak at 944.485 nm and a secondary peak at 657.995 nm).

For the 25-band images, we applied per-pixel normalization (PN) in the preprocessing stage, which involves dividing each of the 25 spectral components by the total sum of all.
By applyin PN, we prioritize the shape of the spectral signatures over variations in mean reflectance values.
Although mean reflectance data can be very useful for class discrimination under constant and controlled illumination conditions, it tends to perform poorly under varying lighting and severely shadowed scenes, as we could observe in the analysis of the HSI-Drive dataset.
Hence, even though this article is not focused on pixel normalization techniques, we observed that while applying PN to the 25-band images provides the best classification results, networks trained with 3-channel and 1-channel images without PN result in generally more accurate predictions than for their normalized counterparts.

\begin{table}[h!]
\centering
\caption{Test IoU metrics with the three models under study.}
\label{tab:metricasIoU}
\resizebox{8cm}{!}{ %
\begin{tabular}{c|c|c|c|c|c|}
\cline{2-6}
 & \textbf{Road} & \textbf{Marks} & \textbf{Veg.} & \textbf{Sky} & \textbf{Other} \\ \hline
\multicolumn{1}{|c|}{\textbf{1-ch.}} & 97.75 & 91.04 & 89.34 & 91.94 & 70.29 \\ \hline
\multicolumn{1}{|c|}{\textbf{3-ch.}} & 98.48 & 93.24 & 91.90 & 92.86 & 77.23 \\ \hline
\multicolumn{1}{|c|}{\textbf{25-ch. (PN)}} & 98.11 & 89.27 & 94.12 & 92.00 & 80.94 \\ \hline
\end{tabular}}
\end{table}

\subsection{Activation and Weight Study for DNN Explainability}
Since none of the saliency methods described in Section \ref{sec:conservativenessEvaluation} provide reliable information for comparing the three models under study (1, 3 and 25 spectral channels), we propose to focus the explainability analysis on features that directly contribute to the computation of the final prediction values in the model.

Firstly, we examined the activations before the final convolution layer ($conv2D\_21$) and the weights and biases of the last convolutional layer ($conv2D\_22$).
The output of $conv2D\_21$ consists of 32 non-negative feature channels with the same size as the input.
Each of the 5 output class nodes has a set of 32 weights and 32 biases.
The first analysis is aimed at determining the amount of relevant feature maps per class, identifying any features that contribute similarly across multiple classes, and finding features with significant positive or negative activations for specific classes.

Due to skip connections in the U-net model \cite{gutierrez2024evaluating}, the second convolutional block ($conv2D\_1$) is directly linked to the final convolutional block.
This helps to preserve spatial information lost during subsequent encoding layers, thus providing particularly meaningful information for this study.
In consequence, secondly we investigated the activations at this initial layer, which usually extract relevant features for edge detection, depending on the number of spectral channels provided at the input.

\begin{figure*}[h!]
  \begin{subfigure}{0.33\textwidth}
    \includegraphics[width=\linewidth]{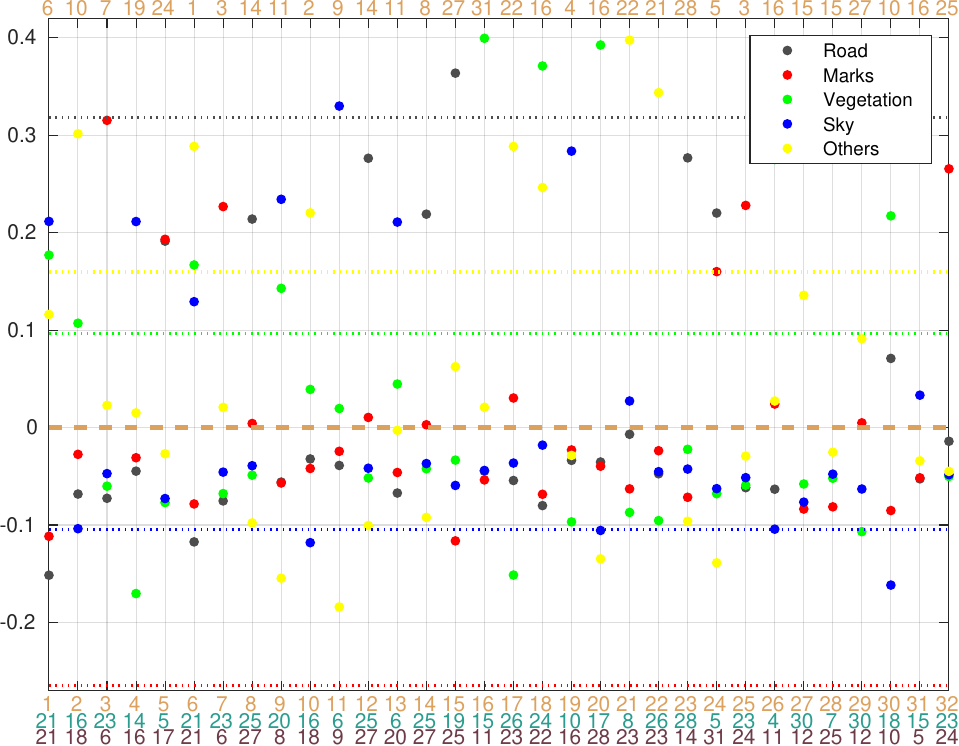}
    \caption{1-channel.}
    \label{fig:1channel}
  \end{subfigure}%
  \hfill
  \begin{subfigure}{0.33\textwidth}
    \includegraphics[width=\linewidth]{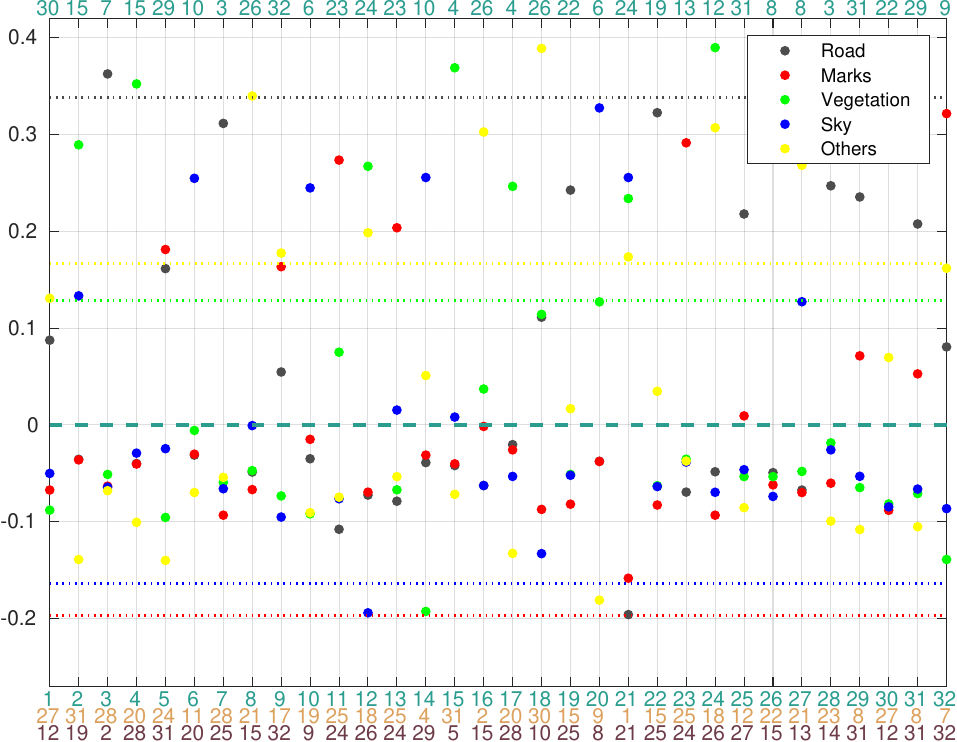}
    \caption{3-channel.}
    \label{fig:3channel}
  \end{subfigure}%
  \hfill
  \begin{subfigure}{0.33\textwidth}
    \includegraphics[width=\linewidth]{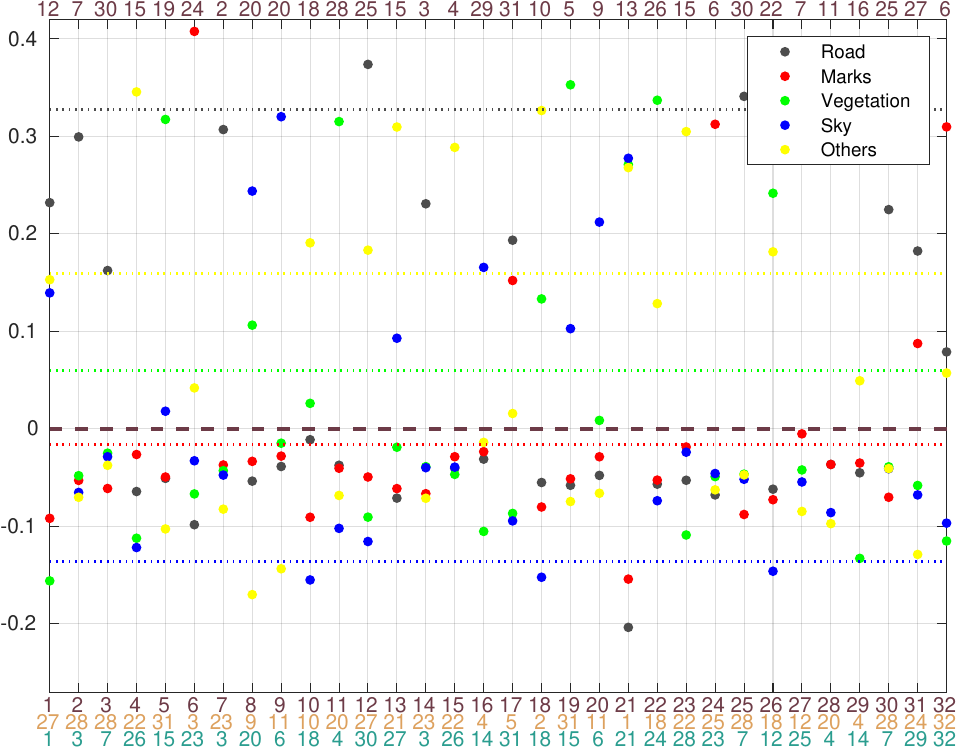}
    \caption{25-channel with PN.}
    \label{fig:25channel}
  \end{subfigure}
  \caption{Weight (dots) and bias (dashed line) values of the 1x1 $conv2D\_22$ convolution layer for each class in the three models.
  Lines below the graph indicate the most correlated weights from other models, while the line above indicates the most correlated weights from the same model: orange (1-channel), turquoise (3-channel), and brown (25-channel with PN).}
  \label{fig:WeightsBiasLastConv}
\end{figure*}

\begin{figure*}[h!]
  \begin{subfigure}{0.33\textwidth}
    \includegraphics[width=5.85cm]{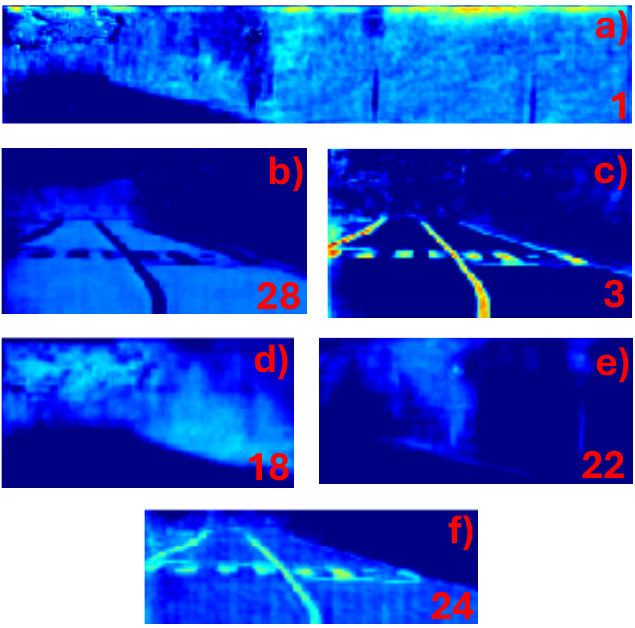}
    \caption{1-channel.}
    \label{fig:1-channelAct21}
  \end{subfigure}%
  \hfill
  \begin{subfigure}{0.33\textwidth}
    \includegraphics[width=5.95cm]{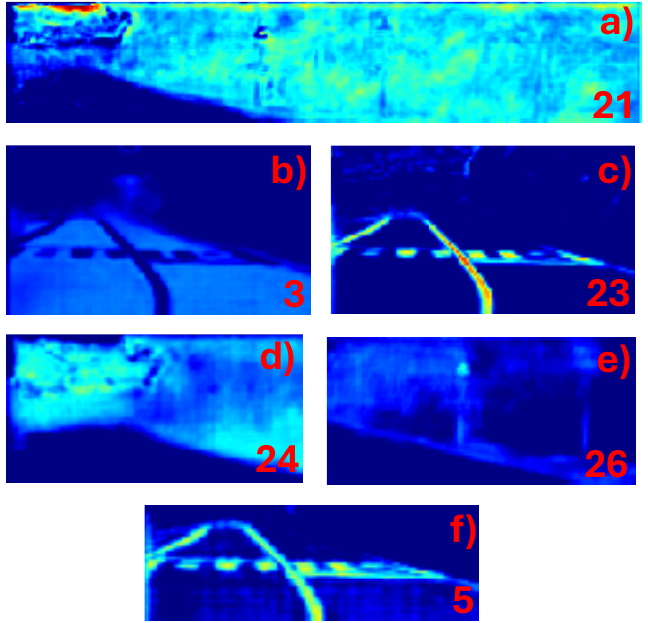}
    \caption{3-channel.}
    \label{fig:3-channelAct21}
  \end{subfigure}%
  \hfill
  \begin{subfigure}{0.33\textwidth}
    \includegraphics[width=5.95cm]{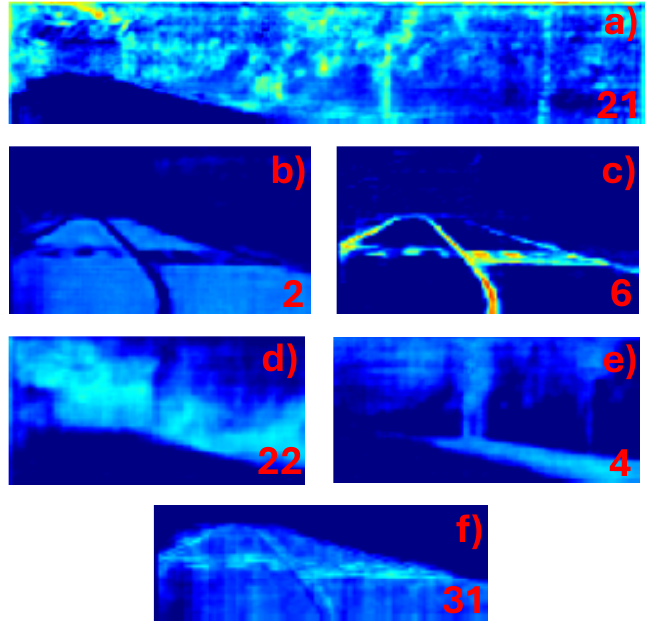}
    \caption{25-channel with PN.}
    \label{fig:25-channelAct21}
  \end{subfigure}
  \caption{Selected outputs/inputs from block $conv2D\_21/22$ of 1-channel (left), 3-channel (center) and 25-channel with PN (right) models.
  Numbers indicate the activation channel.
  Letters indicate highly correlated channels as shown in Fig. \ref{fig:WeightsBiasLastConv}.
  a) Common activation for Veg., Sky and Others.
  b) Strong activation for Road.
  c) Strong activation for Marks.
  d) Homogeneity of Veg. under strong light contrast.
  e) Strong activation for Others.
  f) Common activation for Road and Marks.}
  \label{fig:activation21}
\end{figure*}

\subsubsection{Activation of $conv2D\_21$}
For a comparative visual analysis, we have selected and cropped six activations from this layer, shown in Fig. \ref{fig:activation21}.
To ensure a fair comparison across the three models, we focused on the three most correlated activations.
This election was made by calculating the correlations between the weight vectors of each channel, as shown on the X-axis of Fig. \ref{fig:WeightsBiasLastConv}.
For example, channel 21 of the HSI model shows the highest correlation with channel 1 of the 1-channel model and channel 21 of the 3-channel model.
Across all models, weights associated with Vegetation (green dot), Sky (blue dot), and Others (yellow dot) hover around 0.2, while weights linked to Road (black dot) and Road Marks (red dot) are approximately -0.15.

\begin{figure*}[h!]
\centering
  \begin{subfigure}{0.33\textwidth}
  \centering
    \includegraphics[width=5.5cm]{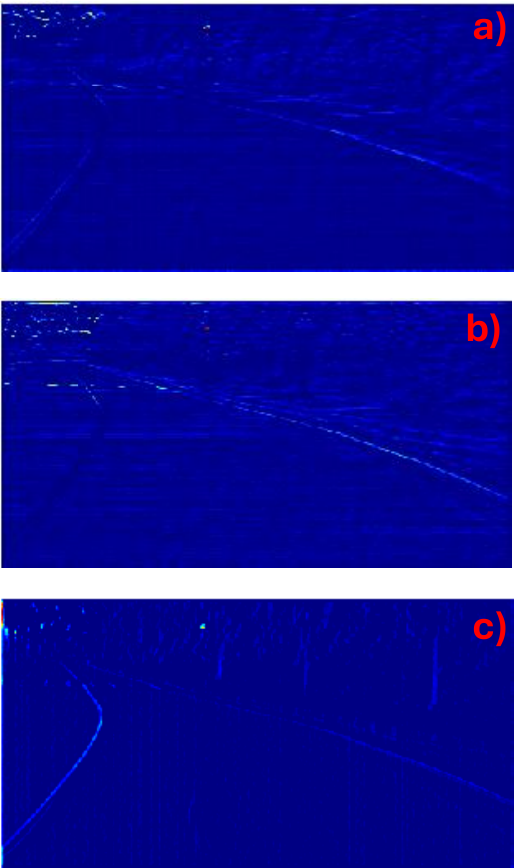}
    \caption{1-channel.}
    \label{fig:1-channelAct1}
  \end{subfigure}%
  \hfill
  \begin{subfigure}{0.33\textwidth}
  \centering
    \includegraphics[width=5.5cm]{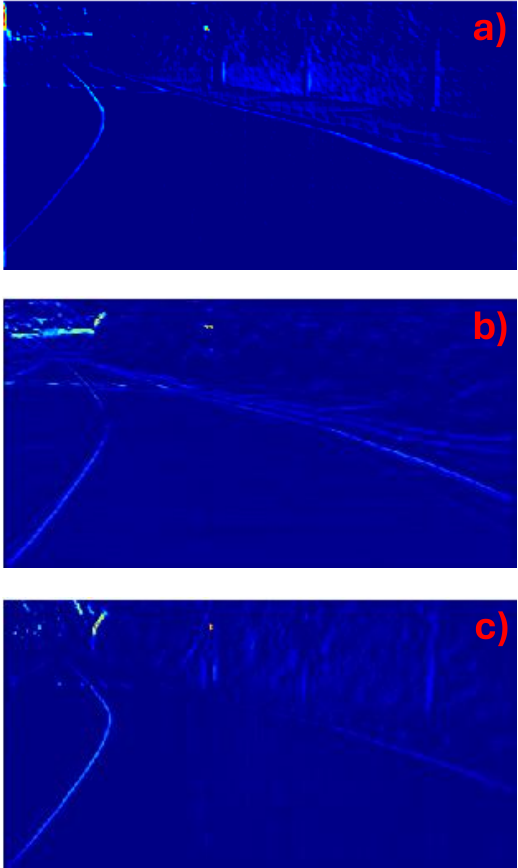}
    \caption{3-channel.}
    \label{fig:3-channelAct1}
  \end{subfigure}%
  \hfill
  \begin{subfigure}{0.33\textwidth}
  \centering
    \includegraphics[width=5.5cm]{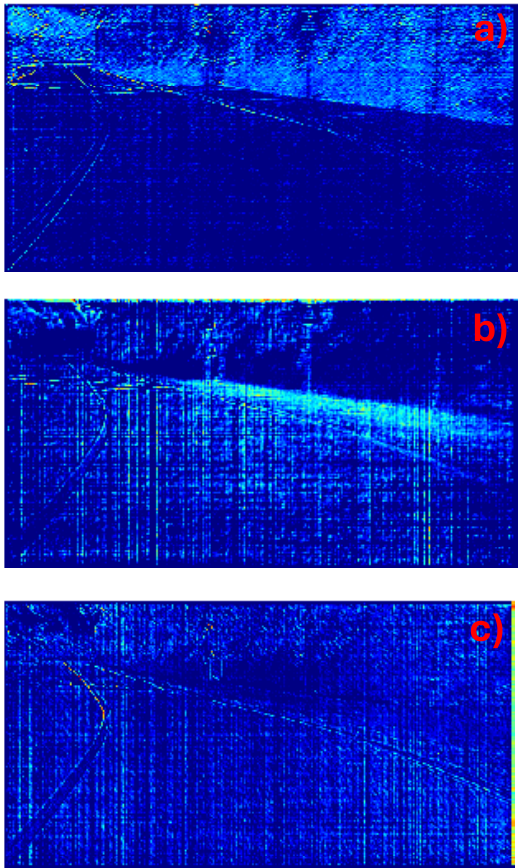}
    \caption{25-channel with PN.}
    \label{fig:25-channelAct1}
  \end{subfigure}
  \caption{Selected outputs from block $conv2D\_1$ of 1-channel (left), 3-channel (center) and 25-channel with PN (right) models.
  a) Border detection in Vegetation.
  b) Border detection in Zebra crossing.
  c) Border detection in curvy Road Mark.}
  \label{fig:activation1}
\end{figure*}

Row $a$ of Fig. \ref{fig:activation21} compares the outputs of three correlated channels that are typically positively activated for Vegetation, Sky, and Other classes, while being negatively activated for Road and Road Marks classes.
It can be observed that the activations in the 25-channel model are less homogeneous in highly diverse areas of the image, indicating an effective utilization of the spectral information.
The advantages of combining HSI information and PN, which aim to homogenize regions partially in sun and shadow, are evident in Figs. \ref{fig:SegmentationExample}a to \ref{fig:SegmentationExample}c and row $d$ from Fig. \ref{fig:activation21}.

In row $b$ of Fig. \ref{fig:activation21}, which shows strong activation for the Road class, we observe certain regions of Vegetation that are erroneously activated in both the 1-channel and, to a lesser extent, 3-channel models.
This highlights a significant drawback of PN: the lack of edge definition, which leads to poorer segmentation of the Road Marks class (Table \ref{tab:metricasIoU}).
An analysis of the channels that are highly activated for the Road Marks class (row $c$ from Fig. \ref{fig:activation21}) and the channels activated for both the Road and Road Marks classes (row $f$ from Fig. \ref{fig:activation21}) reveals this issue further.
While the borders are less sharp for the 25-band model, it is noteworthy that there are no erroneous activations outside the pavement in row $c$ from Fig. \ref{fig:activation21}, a problem observed in the 1-channel and 3-channel models.
Another significant finding in Fig. \ref{fig:WeightsBiasLastConv} is that the number of channels positively activated for Road Marks (with a weight of more than 0.1) is lower for the the 25-channel model with PN, which is compatible with its relative lack of sharpness in the edge detection.

Finally, regarding the separation between the sidewalk (Others) and Vegetation, which is incorrectly classified in the 1-channel case and poorly in the 3-channel case (with the sidewalk classified as Vegetation, as in Fig. \ref{fig:SegmentationExample}a and Fig. \ref{fig:SegmentationExample}b), we see that even in the strongest activation for the Others class, there are no indications of correct sidewalk identification (row $e$ from Fig. \ref{fig:activation21}).
Nevertheless, this demonstrates how normalization diminishes the impact of illumination differences in segmentation.
In this final activation, the traffic light is hardly activated, contrasting with its activation in the 1-channel and especially the 3-channel cases.

\subsubsection{Activation of $conv2D\_1$}
In this case, we used a different criterion to select the channels to display, since the convolutional kernels from layer $conv2D\_2$ are 3x3 instead of 1x1.
We examined the 32 activations from the three models and identified the three most significant ones that exhibit the greatest similarity as shown in Fig. \ref{fig:activation1}.

Due to PN normalization, as seen in Fig. \ref{fig:activation1}c, the activations are noisier and more heterogeneous, leading to defective edge detection of road lines (such as curvy lines and zebra crossings) compared to the 1-channel case, Fig. \ref{fig:activation1}a, and the 3-channel case, Fig. \ref{fig:activation1}b.
However, this also benefits the 25-channel PN model, enabling better separation between the Vegetation and Others classes from the early layers --an effect that is imperceptible in the 1-channel and 3-channel cases.
Additionally, this approach ensures more homogeneous activation in vegetation areas that are under both sunlight and shadow, in contrast to the edge detection presented in Fig. \ref{fig:activation1}b for the 3-channel case.

\section{Discussion and future work}\label{sec:discussion}
As shown in this article, saliency methods show significant limitations for the explainability of DNNs for image segmentation tasks.
We reviewed various tests and invariants proposed in the literature, including network and data randomization, class activation mapping conservativeness, and adversarial examples, to assess the reliability of these methods.

In particular, by extending the definition of conservativeness from image classification to image segmentation, we show that CAM-based methods, which differ in their approaches to obtaining $\alpha_{k}^{c}$, do not meet the conservativeness criteria.
This insight suggests that the utility of these methods for precise model interpretation is questionable.
Furthermore, we observed that most proposed saliency methods prioritize visually appealing representations over accurate depictions of the actual inference processes in the models.

Given these limitations, and with the aim of exploring the advantages of using HSI in segmentation tasks, we propose to perform explainability analysis in segmentation FCNs by analyzing those feature extraction layers that directly contribute to the output predictions.
Although limited, we think this approach assures a higher level of reliability in the analysis.
Specifically, we analyzed activations from the penultimate layer to identify the most strongly triggered activations for each class and from the second layer to evaluate the edge detection process in segmentation tasks.
Our findings indicate that models utilizing a greater number of spectral channels can enhance segmentation performance for classes with high intraclass variability.

This 25-channel model was trained using PN, which favors information provided by spectral signatures rather than general reflectance contrasts.
However, PN can lead to less accurate edge detection, particularly for classes that more strongly rely on reflectance differences.
This is particularly evident in the analysis of the second-layer activations, where, unlike in the 1-channel and 3-channel models, the PN method activates many regions that do not correspond to theoretical edges.
However, the model demonstrates greater robustness to changes in illumination, as evidenced by consistent segmentation of regions under both sunlight and shadow.

These insights highlight the need for ongoing research to develop innovative strategies that effectively leverage the strengths of spectral information while ensuring robust model performance across varying lighting conditions.
Moreover, there is a critical need for better interpretability models specifically tailored for semantic segmentation networks to enhance understanding and trust in these systems.
Future work should focus on creating models that balance the richness of data provided by HSI with the ability to generalize effectively, thereby improving the accuracy and reliability of segmentation outcomes in practical applications, particularly for autonomous driving systems.

\bibliographystyle{IEEEbib}
\bibliography{references} 

\end{document}